\documentclass[aps,pre,twocolumn,superscriptaddress,nofootinbib]{revtex4-1} 
\usepackage{graphicx} 
\usepackage{dcolumn}  
\usepackage{bm}    
\usepackage{amssymb}  
\usepackage{amsmath}  
\usepackage{extarrows}
\usepackage{tikz}
\usepackage[caption=false]{subfig}

\newcommand{\matr}[1]{{#1}}
\newcommand{\bra}[1]{{\left\langle #1 \right|}}
\newcommand{\ket}[1]{{\left| #1 \right\rangle}}
\newcommand{\braket}[2]{{\left\langle #1 \,| #2\right\rangle}}

\newcommand{\hide}[1]{}
\renewcommand{\ao}{\hat{a}}
\renewcommand{\aa}{\hat{a}^\dag}
\newcommand{\bo}{\hat{b}}
\newcommand{\ba}{\hat{b}^\dag}

\newcommand{\no}{\hat{n}}
\newcommand{\ra}{\rangle}
\newcommand{\Ho}{H}
\newcommand{\vo}{\hat{v}}
\newcommand{\la}{\langle}

\newcommand{\refeq}[1]{(\ref{#1})}
\newcommand{\be}{\begin{equation}}
\newcommand{\ee}{\end{equation}}
\newcommand{\bes}{\begin{eqnarray}}
\newcommand{\ees}{\end{eqnarray}}
\begin{document}

\title{On the number of Bose-selected modes in driven-dissipative ideal Bose gases}
\author{Alexander~Schnell}
\email[Electronic address: ]{schnell@pks.mpg.de}
\affiliation{Max-Planck-Institut f{\"u}r Physik komplexer Systeme, N{\"o}thnitzer Stra\ss e 38, 01187 Dresden, Germany}
\author{Roland~Ketzmerick}
\affiliation{Max-Planck-Institut f{\"u}r Physik komplexer Systeme, N{\"o}thnitzer Stra\ss e 38, 01187 Dresden, Germany}
\affiliation{Technische Universit{\"a}t Dresden, Institut f{\"u}r Theoretische Physik and Center for Dynamics, 01062 Dresden, Germany}
\author{Andr{\'e}~Eckardt}
\affiliation{Max-Planck-Institut f{\"u}r Physik komplexer Systeme, N{\"o}thnitzer Stra\ss e 38, 01187 Dresden, Germany}

\date{\today}

\begin{abstract}
In an ideal Bose gas that is driven into a steady state far from thermal equilibrium, a generalized form of Bose condensation
can occur. Namely, the single-particle states unambiguously separate
into two groups: the group of Bose-selected states, whose occupations increase linearly with the total particle number, and the group of all other states whose occupations saturate  [Phys.~Rev.~Lett.~\textbf{111}, 240405 (2013)].
However, so far very little is known
about how the number of Bose-selected states depends on the properties of the system and its coupling to the environment.
The answer to this question is crucial since systems hosting a single, a few, or an extensive number of Bose-selected states
will show rather different behavior. While in the former two scenarios each selected mode acquires a macroscopic occupation,
corresponding to (fragmented) Bose condensation, the latter case rather bears resemblance to a high-temperature state of matter.
In this paper, we systematically investigate the number of Bose-selected states, considering different classes of the rate matrices
that characterize the driven-dissipative ideal Bose gases in the limit of weak system--bath coupling.
These include rate matrices with continuum limit, rate matrices of chaotic driven systems, random rate matrices,
and rate matrices resulting from thermal baths that couple to a few observables only.
\end{abstract}

\maketitle

\section{Introduction}

Other than in thermodynamic equilibrium, the density matrix of
a gas that is driven into a nonequilibrium steady state does
not follow directly from a simple principle like entropy maximization.
It depends on the microscopic details of the system, the bath, and the system--bath coupling.
Lately, a high interest into out-of-equilibrium steady states
which remember the initial condition of an isolated system has developed.
For example, such steady states were observed in many-body localized systems
\cite{BaskoAleinerAltshuler05, NandkishoreHuse15,SchreiberEtAl15, SmithEtAl16,
ChoiEtAl16}, including discrete time crystals
\cite{KhemaniEtAl16, vonKeyserlingkEtAl16, ElseBauerNayak16, ZhangEtAl17,
ChoiEtAl17} that occur in many-body localized time-periodically driven (Floquet) systems.
Also nonequilibrium steady states (NESS) of
driven-dissipative many-body systems have generated much attention. This includes
open Floquet systems \cite{TsujiEtAl09, VorbergEtAl13, VorbergEtAl15,
FoaTorresEtAl14, SeetharamEtAl15, DehghaniEtAl15, GoldsteinEtAl15, IadecolaEtAl15,
ShiraiEtAl16, QinHofstetter17, ChongEtAl2017},
state engineering via dissipation \cite{DiehlEtAl08, LetscherEtAl17,LabouvieEtAl16, SchnellEtAl17}
 and photonic systems \cite{DengEtAl2010,CarusottoCiuti13}, where
 coherent phenomena
in some limits may be related to lasing but in other limits rather
to Bose condensation \cite{SzymanskaEtAl06, KirtonEtAl13, ChiocchettaEtAl14, KlaersEtAl10, ByrnesKimYamamoto14, LeymannEtAl17}.
In this context, optical microcavitities offer great freedom for designing system and dissipative environment, which allows for tailoring the
coherent emission of multiple modes \cite{TuereciEtAl06, Cao03} or to control a switching between emission
of two (or more) different modes \cite{SondermannEtAl03, MarconiEtAl16, GeEtAl16, LeymannEtAl17}.

In this paper, we are focusing on driven-dissipative ideal gases of \(N\) noninteracting bosons
that exchange energy with the environment but no particles. They can be driven out of equilibrium,
e.g. by periodic driving in combination with the coupling to a heat bath or by coupling the system to two
heat baths of different temperature. 
In such setups the ideal gas will relax to a nonequilibrium steady state (NESS) which is characterized by 
a finite heat current through the system.
It is an interesting question, whether (and if yes when and in which form) a
 system can show Bose condensation (or other forms of ordering) under such nonequilibrium conditions.  
 Since this NESS does not follow from thermodynamic principles it is not
obvious whether such a state will feature Bose condensation or not. 
It was observed in Ref.~\cite{VorbergEtAl13} that in the quantum degenerate limit of large densities
the single-particle states split into two groups; 
the \emph{Bose-selected states}, whose occupations increase linearly with the total particle number,
 much like for the ground state in thermal equilibrium, 
while the occupations of all other states saturate.

However, so far very little is known about the factors that determine the number of Bose-selected states. 
The answer to this question is crucial since systems hosting a single, a few, or an extensive number of Bose-selected 
states will show rather different behavior. 
While in the former two scenarios each selected mode acquires a macroscopic occupation, corresponding to (fragmented)
Bose condensation, the latter case rather bears resemblance to a high-temperature state of matter. 
Moreover, inducing transitions between several condensate modes can be a very efficient mechanism 
to exchange energy with the environment, which is not present in systems hosting a single condensate only \cite{VorbergEtAl13}.

In this paper, we investigate how the number of Bose-selected states depends on the properties 
of the system and its coupling to the environment. To this end, we study  several different scenarios, which are reflected in different
forms of the rate matrices that characterize the driven-dissipative ideal Bose gases in the limit of
weak system-bath coupling. These include rate matrices with continuum limit, rate matrices of
chaotic driven systems, random rate matrices, and rates matrices resulting from thermal baths that
couple to a few observables only.

\section{Driven-dissipative ideal Bose gas and Bose selection}

In the limit of weak system--bath coupling, the system will approach a nonequilibrium steady state \(\varrho_S\) that is diagonal in the
eigenstates \(i\) of the Hamiltonian for an autonomous (i.e.\ non-driven) system or in the Floquet states \(i\) for a time-periodically
 driven system \cite{BreuerPetruccione, KohlerEtAl97, VorbergEtAl15}.
 The mean occupations of these states obey the equation of motion \cite{VorbergEtAl13}
\be
\partial_t \la \no_i \ra =	 \sum_{j} R_{ij} \la(\no_{i} +1) \no_{j} \ra - R_{ji} \la(\no_{j} +1) \no_{i} \ra =0.
	\label{eq:eom-ni}
\ee
The rate for a boson to jump from single-particle level~\(j\) to \(i\) is given by the single-particle rate \(R_{ij}\)
multiplied by the bosonic enhancement factor \((\no_{i} +1)\) which manifests that bosons favor to ``jump''
into states that already have large occupation.

It was pointed out in Refs.~\cite{VorbergEtAl13, VorbergEtAl15} that
a generalization of Bose condensation is also observed in the NESS, called Bose selection.
Here, a whole group of an odd number of single-particle states, the {Bose-selected} states, 
can acquire large occupation.
As we briefly recapitulate in this section, these selected states are only determined by the rate asymmetry matrix
\be
	A_{ij} = R_{ij} - R_{ji}.
\ee

We assume that the gas may exchange heat with an environment of one or more thermal phonon baths.
A single bath is described as a collection of harmonic oscillators \(\Ho_\mathrm{B} = \sum_\alpha \hbar \omega_\alpha \ba_\alpha \bo\)
which are in thermal equilibrium.
The corresponding system--bath coupling operator reads \(\Ho_\mathrm{SB} = \gamma \vo \sum_\alpha c_\alpha (\ba_\alpha + \bo)\)
with dimensionless system coupling operator \(\vo\), coefficients \(c_\alpha\) and coupling strength~\(\gamma\).
Throughout the paper we assume that the baths are Markovian.
Thus, the single-particle rate for the autonomous system,
\be
 R_{ij} =\frac{2 \pi \gamma^2}{\hbar} \vert \bra{i} \vo \ket{j} \vert^2 g(\varepsilon_i - \varepsilon_j),
\ee
is of golden rule type. Here, \(\varepsilon_i\) is the energy of single-particle eigenstate \(\ket{i}\). It enters the bath-correlation function,
\be
 g(\varepsilon) = \frac{J(\varepsilon)}{e^{\varepsilon/T}-1},
\ee
where \(T\) is the temperature (measured in units of energy, \(k_B=1\)) of the bath and \(J\) is its spectral function
\(J(\varepsilon) = \sum_\alpha c_\alpha^2 \left[ \delta(\varepsilon-\hbar \omega_\alpha) - \delta(\varepsilon+\hbar \omega_\alpha)\right]\).
In the following we will consider ohmic baths with a continuum of modes \(\alpha\) and spectral function \(J(\varepsilon) \propto \varepsilon\).

A nonequilibrium situation is found when e.g.~the system is coupled  to
multiple baths at different temperatures, where the
total rate is given by the sum of the rates \(R^{(b)}_{ij}\) corresponding to the
 individual bath \(b\), 
 \be
 	R_{ij} = \sum_b R^{(b)}_{ij}.
\ee
Another possible scenario are time-periodically driven systems coupled to a heat bath. In this case the rates read
\be
 R_{ij} =\frac{2 \pi \gamma^2}{\hbar} \sum_{m=-\infty}^{\infty} \vert v_{ij}(m)\vert^2 g(\varepsilon_i - \varepsilon_j-m \hbar \omega)
\ee
with \(v_{ij}(m) =\frac{\omega}{2\pi} \int_0^{2\pi/\omega} e^{\mathrm{i} m\omega t}\bra{i(t)} \vo \ket{j(t)} \mathrm{d}t\),
 driving frequency \(\omega\), Floquet states \(\ket{i(t)}\), and corresponding qua\-si\-en\-er\-gies \(\varepsilon_i\).

\begin{figure}[b]
	\includegraphics{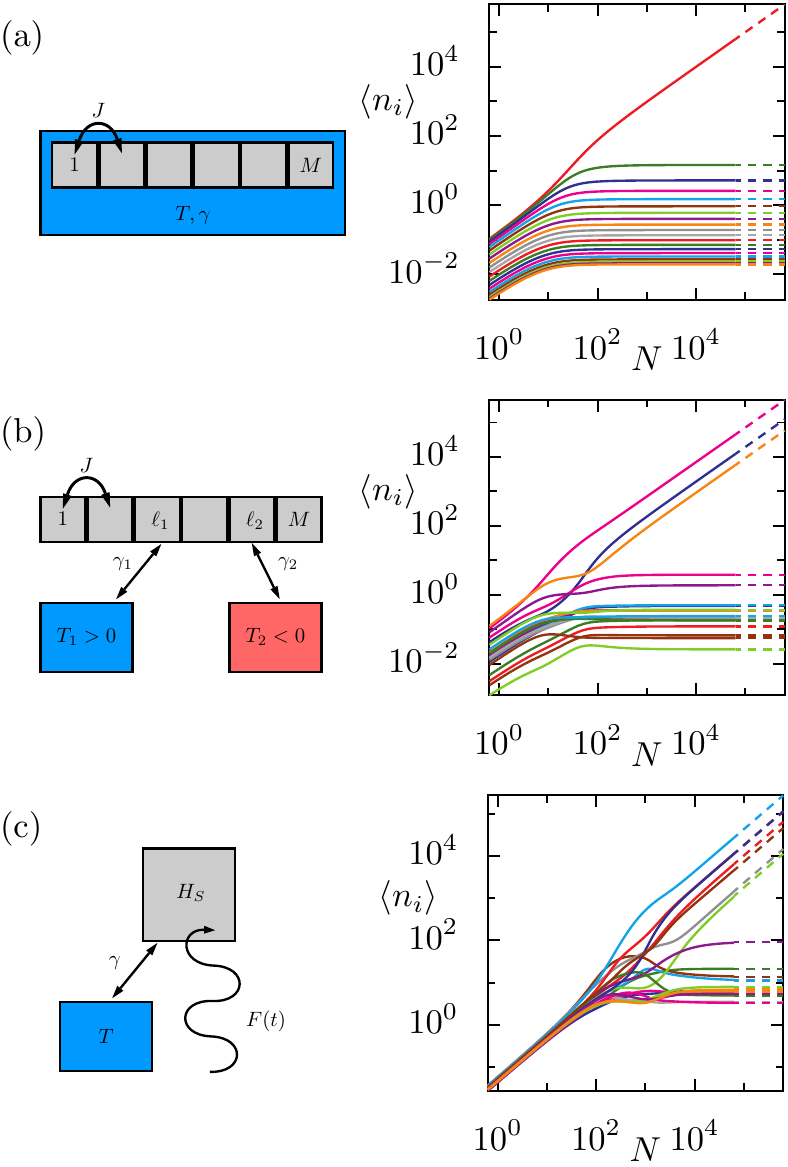}
	\caption{Mean occupations \(\langle \no_i \rangle\) of the single-particle eigenstates~\(i\)
	as a function of the total particle number \(N\) for the nonequilibrium steady state of 
	an ideal Bose gas. Solid lines are from mean field theory, Eq.~\eqref{eq:mf},
	dashed lines from the asymptotic theory. (a,b)~The system is
	 a tight-binding chain with \(M=20\) sites, tunneling parameter~\(J\), that is
	(a) in thermal equilibrium, coupled to a bath at temperature \(T=J\) and (b) in a nonequilibrium steady state,
	coupled to a bath at temperature \(T_1=J\)
	to the occupation number operator at site \(\ell_1 =3\) and a second bath
	 with temperature \(T_2=-0.2J\) at site \(\ell_2=5\) with equal coupling strength \(\gamma_1=\gamma_2\).
	(c) Nonequilibrium steady state of a fully chaotic Floquet system (frequency \(\omega\))
	with \(M=20\) modes, the quantum kicked rotor with
	kicking strength \(K=10\), coupled to a single bath of temperature \(T=\hbar\omega\).}
	\label{fig:main}
\end{figure}

The equation of motion \eqref{eq:eom-ni} for the mean occupations
depends also on the two-particle density--density correlations,
the equations of which depend in turn on three-particle correlations and so on.
In this way it establishes a hierarchy, which 
in the following will be truncated already at the single-particle level by employing the mean-field
decomposition \(\la \no_i \no_j \ra \approx \la \no_i \ra \la \no_j\ra\), \(i\neq j\).
As a result, the steady state occupations, \(\partial_t \la \no_i \ra = 0\), follow from the nonlinear equations of motion
\be
	0 = \sum_{j} A_{ij} \la \no_{j} \ra \la \no_{j} \ra + R_{ij} \la \no_{j} \ra - R_{ji} \la \no_{i} \ra.
	\label{eq:mf}
\ee
Here we have used the rate asymmetries
\(A_{ij}\),
which for the rates of a single bath read
\be
	A^{(b)}_{ij} = \frac{2 \pi \gamma^2}{\hbar} \vert \bra{i} \vo \ket{j} \vert^2 J(\varepsilon_j - \varepsilon_i),
	\label{eq:asym-1b}
\ee
so that they are independent of temperature.
Note that it has been shown by comparison to quasi-exact Monte-Carlo simulations \cite{VorbergEtAl15}
that the mean-field equation \eqref{eq:mf} yields excellent predictions for the mean occupations~\(\la \no_i \ra\) for a broad range of
models\footnote{A possible reason for this good agreement has been pointed out recently \cite{langen2015experimental,lange2017pumping};
a driven system with a set of observables \(\lbrace \hat{A}_i \rbrace\)
which are approximately conserved quantities will relax towards a steady state that
is well described by a generalized Gibbs ensemble
\(
\varrho_\mathrm{GGE} = Z_\mathrm{GGE}^{-1}\exp\left(-\sum_i \lambda_i \hat{A}_i\right).
\)
Due to the weak-coupling limit that we assume, the occupations~\(\no_i\) of the
system's single-particle states~\(i\) are almost conserved, so in our context this set is given by~\(\lbrace \no_i \rbrace\).
For a state \(\varrho_\mathrm{GGE} \propto \exp\left(-\sum_i \lambda_i \hat{n}_i\right)\), the mean-field
decomposition is exact, \(\la \no_i \no_j \ra = \la \no_i \ra \la \no_j\ra\), \(i \neq j\).}.

The solid lines in Fig.~\ref{fig:main} show the steady-state solutions of Eq.~\eqref{eq:mf} as
a function of the total particle number \(N = \sum_i \la \no_i \ra\)
for three different scenarios:
Fig.~\ref{fig:main}(a) shows occupations for a tight-binding chain of \(M=20\) sites,
coupled to one heat bath only; therefore, the steady state is thermal.
Figure~\ref{fig:main}(b) shows occupations for the same chain, but additionally in contact also with a second, population inverted heat bath
described by a negative temperature \(T_2 < 0\). Note that negative temperatures
have been realized, e.g., in atomic quantum gases by preparing a state at the upper edge of a Bloch band~\cite{Braun52}.
Figure~\ref{fig:main}(c) shows occupations for a time-periodically driven quantum kicked rotor with \(M=20\) Floquet states
coupled to a single bath. The system is in a regime, where the
corresponding classical system, the Chirikov standard map~\cite{Chirikov1979}, is known to be chaotic.

For small total particle number \(N\),
the bosons behave classically and the occupations \(\la \no_i \ra\) are given by the single-particle probabilities \(p_i^{\mathrm{sp}}\)
to occupy the state~\(i\) scaled linearly with particle number \(N\), \(\la \no_i\ra \simeq p_i^{\mathrm{sp}} N\).
However, at large total particle numbers the bosonic quantum statistics makes itself felt.
As a result, we observe Bose selection \cite{VorbergEtAl13}: the occupations of some of the
states saturate, while all additional particles gather in a set \(S\) of selected states whose occupations grow linearly with \(N\).
 In equilibrium, Fig.~\ref{fig:main}(a),
this corresponds to Bose condensation in the ground state. 
Away from equilibrium several states can be Bose selected. 

From Figs.~\ref{fig:main}(a)-\ref{fig:main}(c) we already observe that the number \(M_S = {\vert S \vert}\) of selected states can range
from only few up to an extensive number, while the former case corresponds to fragmented Bose condensation,
since each of the selected state acquires a macroscopic occupation in the limit \(N\to\infty\), the
latter case does not correspond to Bose condensation, since none of the selected states will acquire a macroscopic occupation.
To be more precise, 
in the thermodynamic limit, \(N,M \rightarrow \infty\), \(N/M=\mathrm{const.}\), there can only be (fragmented) condensation if the
number \(M_S\) of Bose-selected states is intensive, i.e. asymptotically independent of \(M\).
However, if there is an extensive number of states, \(M_S \propto M\), the system will behave effectively
classically also in the ultra degenerate limit.
So far, however, very little is known about how \(M_S\) depends on the properties of the system and
in particular on the form of rates. The main goal of this paper is to obtain a better understanding
of how the number of selected states is determined by the properties of the rate matrix.

Before we begin with our analysis,
let us briefly review the equations that determine the set of selected states and what so far has been known about their number.
It has been shown that generally the number \(M_S\) of these selected states is odd.
The starting point for determining the set of selected states is an asymptotic expansion (dashed lines in Fig.~\ref{fig:main})
of the mean occupations in the limit of large occupation, \(N \gg 1\).
For the selected states, \(i \in S\), Eq.~\eqref{eq:mf} yields in this limit \cite{VorbergEtAl13, VorbergEtAl15}
\be
	0 = \sum_{j \in S} A_{ij} \eta_{j}, \forall i \in S,
	\label{eq:N-sel}
\ee
where \(\eta_{i}\) are the leading order occupations \(\la \no_i \ra = \eta_{i} N + \mathcal{O}(N^0)\).
Note that for the nonselected states, these contributions vanish, so it must hold \(\eta_{i} = 0, i\notin S\).
The vector \(\eta_i, i\in S\), is thus a nontrivial vector in the kernel of the skew-symmetric
matrix \({A}_S = \lbrace A_{ij} \vert i,j \in S \rbrace\).
The leading order of the occupations of the nonselected states, \(i \notin S\), is then given by
\be
	 \la \no_{i} \ra = - \frac{\sum_{j\in S} R_{ij} \eta_{j}}{\sum_{j \in S} A_{ij} \eta_{j} } + \mathcal{O}(N^{-1}).
	\label{eq:N-nonsel}
\ee
The physical condition of having positive occupations, \(\la \no_{i} \ra > 0\) for both the selected and the nonselected states,
has been shown to uniquely determine the set of the selected states \cite{VorbergEtAl15}.
Using equations \eqref{eq:N-sel} and \eqref{eq:N-nonsel} this condition can be cast into the simple form
\be
	\mu = \matr{A} \eta \text{ with } \left\lbrace
	\begin{array}{c}
	\mu_{i} = 0, \eta_{i} > 0, \text{ for }  i\in S\\
	\mu_{i} < 0, \eta_{i} = 0, \text{ for } i\notin S.
	\end{array} \right.
	\label{eq:MS-problem}
\ee

From this condition a few simple statements about the number of selected states have been drawn already.
First, we know that \(M_S\) is generically odd, since without fine tuning the skew-symmetric \(A_S\) possesses
a non-trivial kernel for an odd number of selected states only.
Second, in case the system possesses a ground-state like state \(k\) defined by
\be
	R_{ki} >R_{ik}, \forall i \neq k,
	\label{eq:ground-state-like}
\ee
 one can immediately see that the problem \eqref{eq:MS-problem} is solved by \(S=\lbrace k\rbrace\),
 i.e.~a single selected state is found.
Finally, for uncorrelated random rates it has been observed numerically that the number of selected states follows
a binomial distribution. Thus, an extensive number of states (on average half of the states) are selected.

However, a general estimate for \(M_S\) is not straight forward.
In the following, we will discuss different scenarios
in which such estimates can be found; First we will study rates
which possess a well-defined continuum limit for \(M \to \infty\).
Second we will consider rates that do not have such a continuum limit,
discussing the two important cases where rates are truly random, and
rates that stem from a chaotic kicked system.
Finally, we will discuss rates that are given by a sum of direct products,
as they are relevant for autonomous systems that couple to the environment via a few observables only.

\section{Rates with continuum limit}
In this section we discuss systems described by rate matrices that have a smooth continuum limit and which are, thus, strongly correlated.
We assume that the quantum numbers \(i \in \left\lbrace 0, ..., M-1\right\rbrace\) can be labeled by a variable
\be
	k_i = \alpha \frac{i}{M}, \text{ with constant } \alpha \in \mathbb{R},
	\label{eq: discretization}
\ee
that becomes continuous in the limit \(M \rightarrow \infty\).
Moreover, we focus on one-dimensional systems whose rate matrices shall become smooth in that limit,
i.e.~that there exists a function \(R(k,q)\), such that
\be
	R_{ij} = R(k_i, k_j) \Delta^2
\ee
with \(\Delta = \alpha/M\). The generalization to higher-dimensional systems described by several continuous
quantum numbers is straight forward.

An example for this situation are the rates for a single bath coupled to site \(\ell\) of the tight-binding chain
described by the Hamiltonian [see Fig.~\ref{fig:main}(b)]
\be
	H_S = -J \sum_{i=1}^{M-2} (\aa_{i+1}\ao_i + \aa_{i}\ao_{i+1}),
	\label{eq:H-tb}
\ee
 where \(J\) is the tunneling constant and \(\ao_i \) is the bosonic annihilation operator at site \(i\).
In this case we find \cite{SchnellEtAl17}
\be
	R(k,q) = \frac{2 \gamma^2}{\pi \hbar} g(\varepsilon(k)-\varepsilon(q)) \sin(k \ell)^2 \sin(q \ell)^2
\ee
with dispersion \(\varepsilon(k) = -2J \cos(k)\) and \(k\)-space sampling with \(i = 1, \dots, M-1\), and \(\alpha = {\pi}\).

\begin{figure}[t]
	\includegraphics[scale=1]{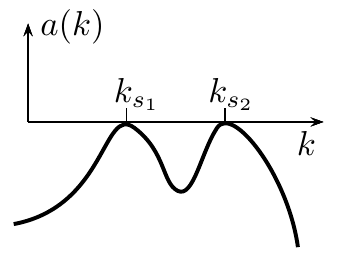}
	\caption{Sketch of the typical behavior of the function \(a(k)\) defined in Eq.~\eqref{eq:func-a-k}.
	The function is globally negative with zeros at the Bose-selected states \(k_s\).}
	\label{fig:sketch-a-k}
\end{figure}

We make the ansatz that there is a discrete set of Bose-selected states \(S = \lbrace k_s \rbrace\), such that
in the asymptotic limit for large densities \(n \rightarrow \infty\) the mean occupation density reads
\be
	\la n(k)\ra = \la n_n(k)\ra + \bar{n}_s \delta(k-k_s),
\ee
where we have introduced the normalization condition
\be
	n = N \Delta = \sum_i  \la n_{i} \ra \Delta \Rightarrow n= \int_0^\alpha \la n(k)\ra \mathrm{d}k.
\ee
Eq.~\refeq{eq:N-sel} then translates to
\be
	 a(k_s)=0, \quad \forall k_s \in S,
	\label{eq:n-sel-cont}
\ee
where we have defined the function
\be
	a(k) = \sum_{s\in S} A(k, k_s) \, \bar{n}_s,
	\label{eq:func-a-k}
\ee
with rate asymmetry function \(A(k,q) = R(k,q) - R(q,k)\).
Note that for smooth rates, the function \(a(k)\) will be smooth as well.
The starting point for our reasoning is the analog of Eq.~\eqref{eq:N-nonsel}, which predicts
that in the continuum limit the asymptotic occupations of all modes \(k \notin S\) read
\begin{equation}
	\la {n}_n(k) \ra = - \frac{\sum_{s\in S} R(k, k_s) \, \bar{n}_s}{a(k)}.
	\label{eq:n-unsel-cont}
\end{equation}
Since here the numerator is strictly non-negative,
the denominator has to be strictly negative
\be
	 a(k) < 0, \quad \forall k \notin S.
	\label{eq:unsel-cont}
\ee

In the following, we assume that \(R\) is two-fold differentiable.
By discussing the rate function in the vicinity of the selected states, we will then be able to restrict the possible selected states to a few
candidates only.
From Eqs.~\eqref{eq:n-sel-cont} and \eqref{eq:unsel-cont} it follows that \(a(k)\) is
negative almost everywhere, however at local maxima it assumes
the value zero, whenever \(k=k_s\), 
see sketch in Fig.~\ref{fig:sketch-a-k}.
This implies both
\begin{align}
	0 = a^\prime(k_s) = \sum_{p \in S} A^{\prime}(k_s, k_p) \ \bar{n}_p,
	\label{eq:deriv-criterion}
	\intertext{and}
	0 > a^{\prime\prime}(k_s) = \sum_{p \in S} A^{\prime\prime}(k_s, k_p) \ \bar{n}_p,
	\label{eq:deriv-2-criterion}
\end{align}
where we have defined \(A^{\prime}(k, q) = \partial_k \, A(k, q)\) and \(A^{\prime\prime}(k, q) = \partial_k^2 \, A(k, q)\).
Note that these local criteria are necessary, but not sufficient for Bose selection in state \(k_s\).
Note also that state space $k$ may have a boundary or not.
For example for the tight-binding chain in Fig.~\ref{fig:main}(b),
we do not impose periodic boundary conditions in real space,
then \(k\) takes values in the interval \(\left[0, \pi \right]\).
Thus the dispersion \(\varepsilon(k)\) is not a periodic function of \(k\) and our state space possesses boundaries at \(0\) and \(\pi\).
At such boundaries the criteria \eqref{eq:deriv-criterion} and \eqref{eq:deriv-2-criterion} do not have to apply,
because here the maxima of \(a(k)\) are no longer characterized by derivatives.

As we will see, the criteria \eqref{eq:deriv-criterion} and \eqref{eq:deriv-2-criterion} strongly constrain the set of selected states.
To illustrate this fact, we will discuss different scenarios in the following subsection.

\begin{figure}[t]
	\includegraphics{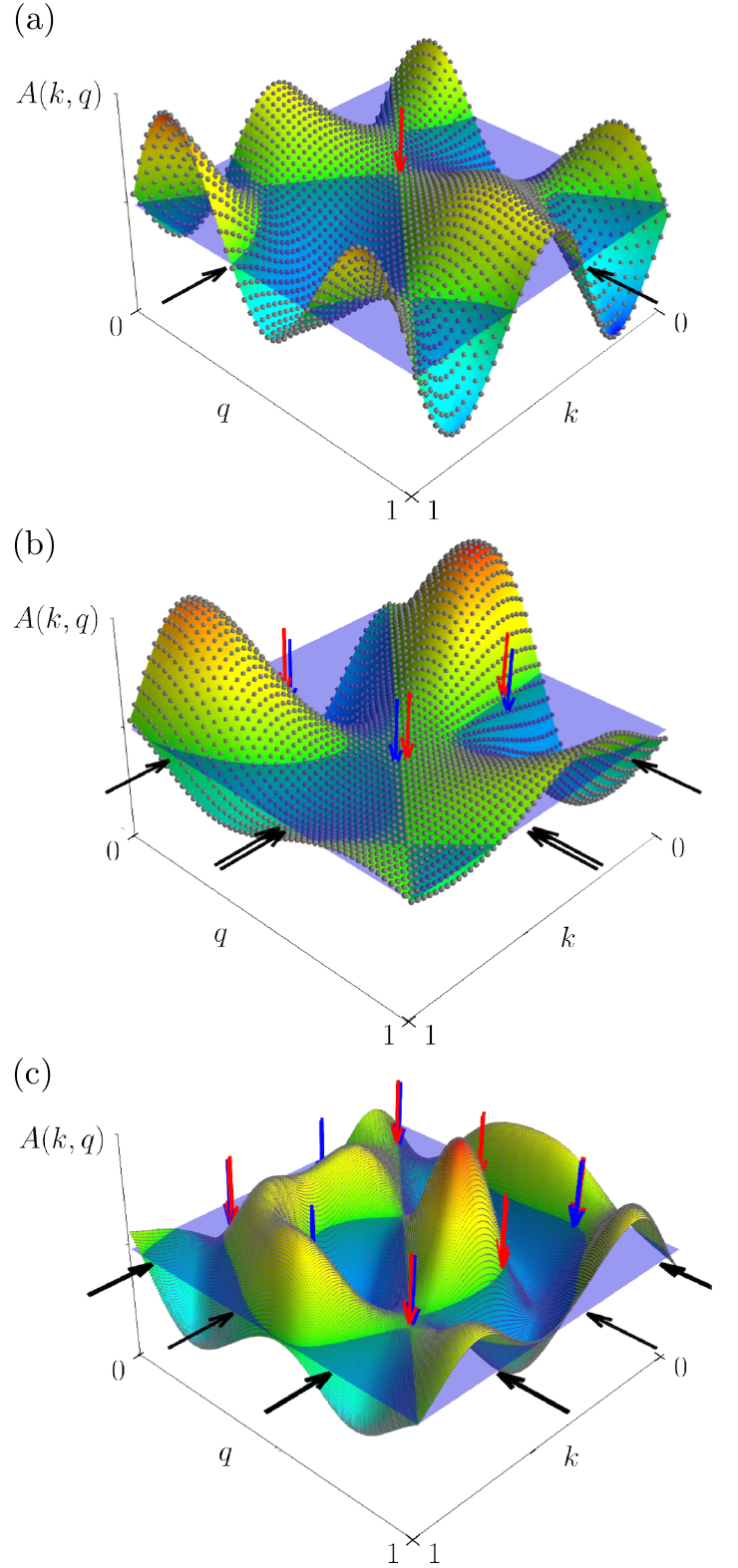}
	\caption{ Gallery of selected states in
	the random wave model, Eq.~\eqref{eq:rand-wave}, for a superposition of \(L=80\) plane waves with wavenumber
	\(\kappa={3\pi}\) (a, b) and \(\kappa={5\pi}\) (c). We discretize \(k\)-space such that \(M=50\) (a, b) and
	 \(M=200\) (c) states exist.
	The smooth rate-asymmetry function \(A(k,q)\)
	is sampled at the grey points.
	The blue plane represents \(A=0\).
	Selected states are marked by black arrows at the \(k\)- {and} \(q\)-axis. We indicate
	the contributing matrix elements \(A(k_s, k_p)\) by red (for \(A\geq0\)) and blue (for \(A < 0\)) arrows.
	For clarity we do not mark diagonal elements \(A(k_s, k_s)\) if \(M_S > 1\).}
	\label{fig:rand-wave-examples}
\end{figure}

\subsection{Different selection scenarios}

If only one state \(k_0\) is selected and does not lie at the boundary of the state space, it follows that
\begin{align}
	0 = \partial_k \, A(k, q) \vert_{(k_0, k_0)}\quad
	\text{and}\quad
	0 > \partial_k^2 \, A(k, q) \vert_{(k_0, k_0)}.
\end{align}
Since \(A\) is skew-symmetric,
\begin{align}
	A(k,q) = - A(q,k), \forall k,q,
\end{align}
we also find
\begin{align}
	0 = \partial_q \, A(k, q) \vert_{(k_0, k_0)}\quad
	\text{and}\quad
	0 < \partial_q^2 \, A(k, q) \vert_{(k_0, k_0)}.
\end{align}
Since also the gradient of \(A\) vanishes at \((k_0, k_0)\), this point must be an extreme point of the asymmetry function \(A\).
Moreover, we find that the mixed derivative at this point must vanish
\begin{align}
	\begin{split}
	&\partial_k \partial_q \, A(k, q) \vert_{(k_0,k_0)} = \partial_q \partial_k \, A(k, q) \vert_{(k_0,k_0)}\\
	& = - \partial_q \partial_k \, A(q, k) \vert_{(k_0,k_0)} = - \partial_k \partial_q \, A(k, q) \vert_{(k_0, k_0)} = 0.
	\end{split}
\end{align}
Here we first used Schwarz' theorem and in the third step renamed the variables.
Thus $k_0$ either corresponds to a saddle point on the diagonal of the rate asymetry matrix 
or it lies at the boundary of state space (if there is one), where both
local criteria do not have to apply.

An example is given by the rate-asymmetry function~\(A\) shown in Fig.~\ref{fig:rand-wave-examples}(a).
Here we mark the position of the selected state by black arrows on the side
and the relevant matrix element \(A(k_0, k_0)\) by a red arrow.
It lies at a saddle point (having the correct curvature) on
the diagonal. Clearly, also the rest of function \(a(k) \propto A(k, k_0)\)
must remain below the blue \(A=0\) plane.

For \(M_S > 1\) the vector of the occupations \(\lbrace \bar{n}_s \rbrace_{s \in S}\) is a homogeneous solution of
Eq.~\eqref{eq:n-sel-cont} and at the same time of Eq.~\eqref{eq:deriv-criterion}.
This provides a strong restriction, since both equations generally will not have a common set of solutions.
Therefore, 
selected states  will generically occur in two special scenarios.

The first scenario is the following. Since the rate asymmetry \(A\)
is continuous we will naturally find zero lines \(A(k, q)=0\)
also away from the diagonal \(A(k, k)=0\).
If the selected states lie at these zero lines, i.e.~\(A(k_s, k_p) =0\),
the occupations \(\lbrace \bar{n}_s \rbrace_{s \in S}\) are only determined by Eq.~\eqref{eq:deriv-criterion}.

Note that since the coefficient matrix \(\left(A^{\prime}(k_s, k_p)\right)_{s,p}\) is not
necessarily skew-symmetric, in this case also an even number of selected states \(k_s\) may occur in the continuum limit.
In the discrete case, an even number $M_S$ of selected states requires 
fine-tuning in the rate matrix \cite{VorbergEtAl15} such that for example some of the $A_{ij}$ vanish.
However, for continuous rate asymmetry functions it is natural to have zero lines,
i.e.~\(A(k_s, k_p)=0\) for $k_s \neq k_p$, such that in the continuous case no fine-tuning is needed to observe an even number $M_S$.
However, if an even number of selected states \(k_s\) occurs in the continuous model, a corresponding discrete system will still feature an
odd number of selected states.
We then typically find pairs of neighboring selected states around at least one\footnote{Around
which of the states \(k_s\) pairs form is, however, not universal and depends on the discrete grid.
If one plots the selected states as a function of discretization parameter \(M\) for example,
one can observe how such pairs jump from one \(k_s\) to the other quite irregularly (not shown).}
of the selected states \(k_s\) of the continuous model [as in the example
in Fig.~\ref{fig:rand-wave-examples}(b)]. To avoid confusion,
 we refer to the number of selected states in the continuous system
as \emph{continuum number of selected states} \(M_S=1,2,3, 4,\dots\).

In Fig.~\ref{fig:rand-wave-examples}(b) we observe Bose {selection at zero lines} where the
continuum selection of \(M_S=2\) states occurs.
The asymptotic states \(k_1,k_2\) lie at a zero line of \(A\). Since in the discrete system
the number of selected modes \(M_S\) is always odd, we find \(M_S=3\) in the discrete system with a pair of neighboring states in
the vicinity of state \(k_2 > k_1\).


Another possible scenario is that the selected states are found such that at $(k_s,k_p)$
the rate asymmetry function possesses saddle points. Then \(A^{\prime}(k_s, k_p) = 0\)
such that the occupations \(\lbrace \bar{n}_s \rbrace_{s\in S}\) are solely determined by Eq.~\eqref{eq:n-sel-cont}.
In this case the corresponding continuum \(M_{S}\)
will be odd.

\begin{figure}[t]
	\includegraphics[scale=0.23]{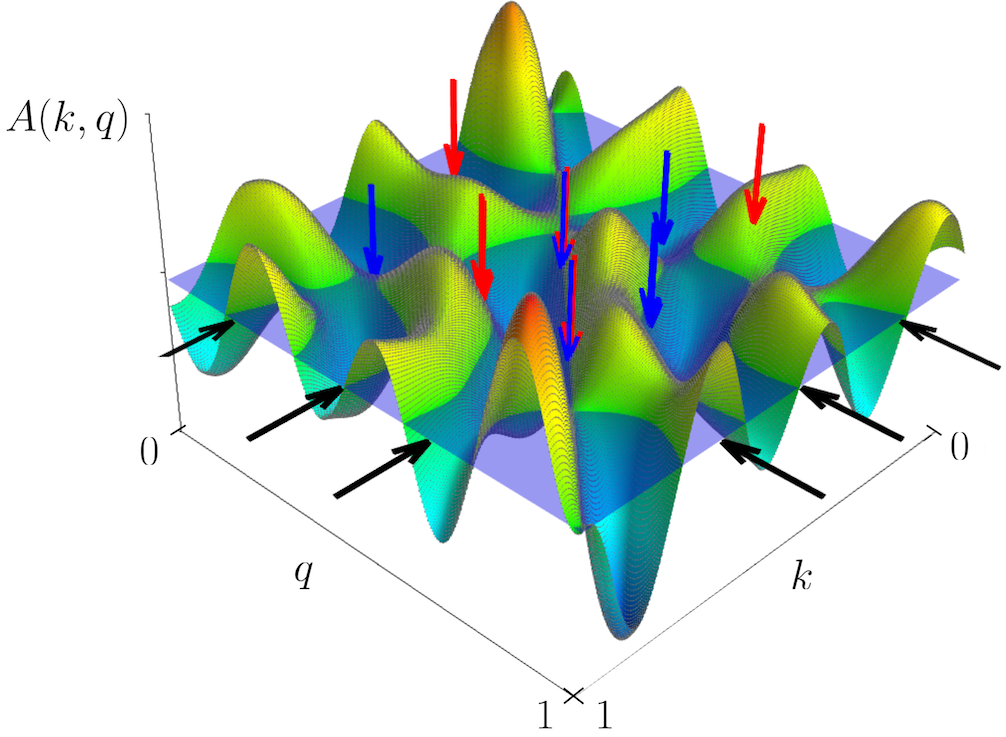}
	\caption{Typical example for Bose selection in a more complex rate matrix.
	Parameters are like in Fig.~\ref{fig:rand-wave-examples}, but $M=300$ and $\kappa=7\pi$.}
	\label{fig:rand-wave-examples-complex}
\end{figure}

However typically neither of the cases -- selection at only zero lines
or only saddle points --  occurs in a ``pure'' form.
This can be seen in the example in Fig.~\ref{fig:rand-wave-examples}(c),
where we observe the {selection at both zero lines and saddle points} with a continnum selection of \(M_S=3\) states.
In this case, some of the relevant points \((k_s, k_p)\) lie on the zero lines of \(A\), and others
have saddle points in the vicinity of the point \((k_s, k_p)\). Here,
\(M_S=5\) states are selected in the discrete system, with pairs at two continuum wave numbers $k$.

But it is only in continuous rate matrices with few oscillations
(like in the examples chosen in  Fig.~\ref{fig:rand-wave-examples}) that we find selected states defined by one of the 
different mechanisms stated above.
If we consider systems with more variation in the rate matrix, like in Fig.~\ref{fig:rand-wave-examples-complex},
 the points $(k_s,k_p)$ are not clearly relatable
to either zero lines nor saddle points anymore. 
Nevertheless, these points are often still found in the vicinity of zero lines and saddle points.



\subsection{Random-wave model}
\label{sec:rand-wave}
\begin{figure}[t]
	\includegraphics{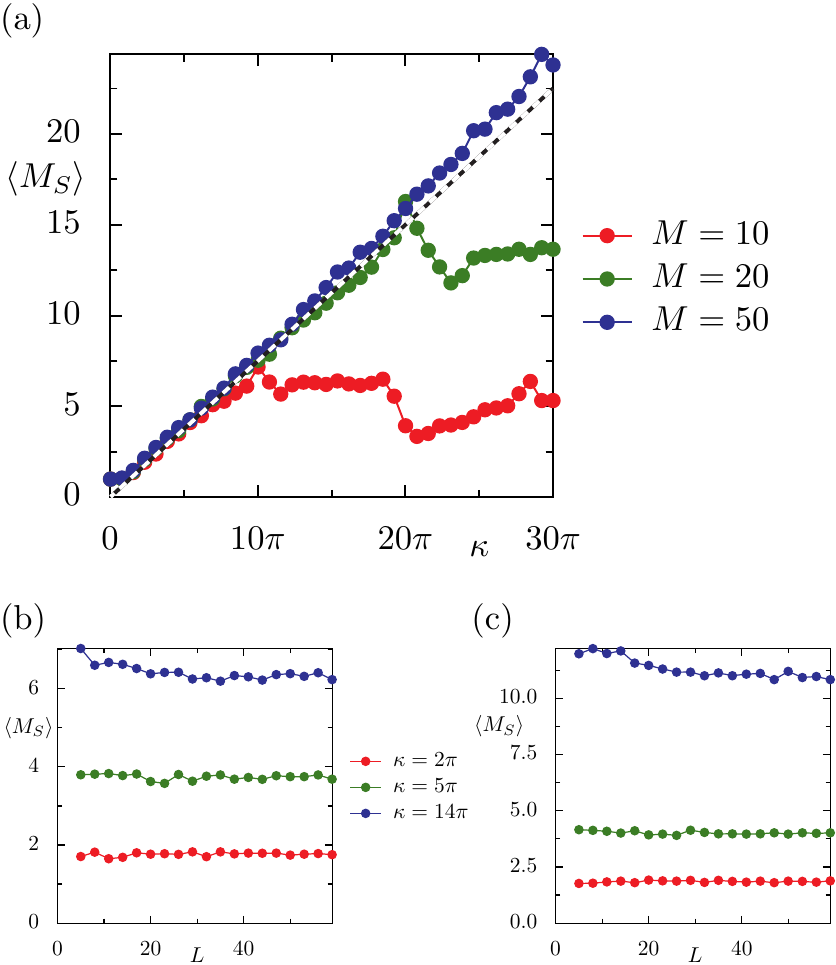}
	\caption{
	Mean number \(\la M_S \ra\) of selected states in the random wave model \eqref{eq:rand-wave}
	for (a) a superposition of \(L=50\) plane waves as a function of the wave-number
	\(\kappa\). The rates \(R(k,q)\) are sampled at \(M=10\), \(20\) and \(50\) discrete states.
	We average over \(500\) realizations of the random rates.
	We see a linear increase with \(\la M_S \ra \approx 0.75 \kappa/\pi\) (dashed line, guide to the eye)
	before it saturates at a value where more than half of the single particle states are Bose selected.
	(b, c) Mean number of selected states as a function of the number \(L\) of components for
	(b) \(M=10\) and (c) \(M=20\) discretization steps
	and 250 realizations.
	We see only a weak dependence on \(L\).}
	\label{fig:rand-wave-func-k}
\end{figure}

The rate functions
 shown in Fig.~\ref{fig:rand-wave-examples}  and \ref{fig:rand-wave-examples-complex} are motivated by a  random wave model for chaotic
 eigenfunctions \cite{berry1977}.
They are defined by
\be
	R(k, q) = \sum^L_{l=1}\ \mathrm{Re}\big\lbrace c_l \exp\left[\mathrm{i} (\kappa_{k,l} k + \kappa_{q,l} q)\right]\big\rbrace + C.
	\label{eq:rand-wave}
\ee
It is a superposition of \(L\) independent plane waves with
\(\kappa_{k,l} = \kappa \sin(\varphi_l)\), \(\kappa_{q,l} = \kappa \cos(\varphi_l)\), fixed absolute value of the
wavenumber \(\kappa\) and uniformly distributed angles \(\varphi_l \in \left[0, 2 \pi \right)\).
Amplitudes \(\left\vert c_l \right\vert\) are drawn from a normal distribution and phases \(\mathrm{arg}(c_l)\)
are distributed uniformly. The global constant \(C\) is chosen such that all rates are non-negative.
Its specific value is irrelevant for the set of the Bose-selected states, which are solely determined
by the rate-asymmetry function \(A(k,q)\) in which \(C\) will drop out.

This model produces rate functions that are smooth and show oscillations on the length
scale \(2\pi / \kappa\) in state space, see e.g.~Fig.~\ref{fig:rand-wave-examples}.
It allows us to investigate the typical behavior of
smooth rates that show variations on such a scale.
We discretize state space via Eq.~\eqref{eq: discretization}
where, to unravel the continuum physics, we choose the discretization length \(\Delta\) to be small with respect to the
oscillation length of the model, \(\Delta=1/M \ll 2\pi/\kappa\).

\label{sec:M_s-func-k}

As we increase the wavenumber \(\kappa\) of the model the structure of \(R\) will become more and more complex, leading
to more and more zero-lines and saddle points in the rate asymmetry function.
Thus, we expect that with the characteristic wavenumber \(\kappa\), which determines the typical oscillation length \(2\pi /\kappa\) of the
smooth rate function \(R\), also the average number of selected modes \(\langle M_S \rangle\) will increase.
This is confirmed by the numerical results shown in Fig.~\ref{fig:rand-wave-func-k}(a).
 The mean number of selected
states increases as soon as \(\kappa\) exceeds the threshold \(\kappa \approx \pi\), where the
wavelength \(2\pi /\kappa\) of the plane waves is about double the system size.
Afterwards, we see a linear increase with \(\la M_S \ra \approx 0.75 \kappa/\pi\) as marked by the
black and white dashed line.
The linear scaling is explained by the above reasoning, since e.g.~the number of extrema (and also the number of zeros)
of \(\sin(\kappa k)\) on the interval \(k \in [0,1]\) is of the order of \(\kappa/\pi\). 
However the origin of the prefactor of about \(0.75\) remains open.

Note that the mean number \(\langle M_S \rangle\) only depends weakly on the number of components \(L\)
that we use in the random wave model as shown in Fig.~\ref{fig:rand-wave-func-k}(b) for
 \(M=10\) discrete states and (c) \(M=20\) and three different values of~\(\kappa\). 

The behavior seen in Fig.~\ref{fig:rand-wave-func-k}(a) clearly suggests that for smooth rates, the number of selected states
is typically on the order of the number of oscillations in the rate asymmetry function \(A\).
Therefore, for fixed smooth \(A\), even if the number of discrete states \(M\) is large, \(M \rightarrow \infty\),
as observed in Fig.~\ref{fig:rand-wave-func-k}, the number \(M_S\) will remain
intensive as it is solely the property of the smooth function \(A\).

We expect a breakdown of the theory for continuous rates as soon as the oscillation length
of the rate function becomes comparable to the discretization length.
For the random rates this breakdown occurs for  \(\kappa \approx M \pi \) as visible in Fig.~\ref{fig:rand-wave-func-k}(a).
Interestingly, the mean number of selected states \(\langle M_S \rangle\)
for the random wave model does not saturate
at \(\langle M_S \rangle = M/2\) as one would expect for truly random rates (see Sec.~\ref{sec:random-rates}).
We observe a first saturation at values that are slightly above \(M/2\): for \(M=20\)
discrete states we find saturation at about \(\langle M_S \rangle = 13\), or for \(M=10\)
we find \(\langle M_S \rangle \approx 6\).\footnote{Note that after this first saturated regime,
for the red line \(M=10\), the number drops again at around \(\kappa = 2 \pi M\).
This is related to the effect of aliasing occuring when the wavelength \(2 \pi / \kappa\) of the random
wave model is smaller than the discretization length \(\Delta = 1/M\).}

\section{Rates without continuum limit}
\label{sec:floquet-chaotic}

\subsection{Uncorrelated random rates}

The single particle rates \(R_{ij}\) that one observes typically for a fully chaotic quantum rotor 
have been shown to roughly follow an exponential distribution \cite{Wustmann10}.
If we suppose that  there are no additional correlations between the rates \(R_{ij}\),
we may draw typical rates from a random realization of the exponential distribution,
where \(p(R_{ij}) = \exp(-\lambda R_{ij})/\lambda\).
Note that the choice of the parameter \(\lambda\) is irrelevant, since it only determines the time scale
on which the system relaxes but not the steady state.

\begin{figure}[t]
	\includegraphics{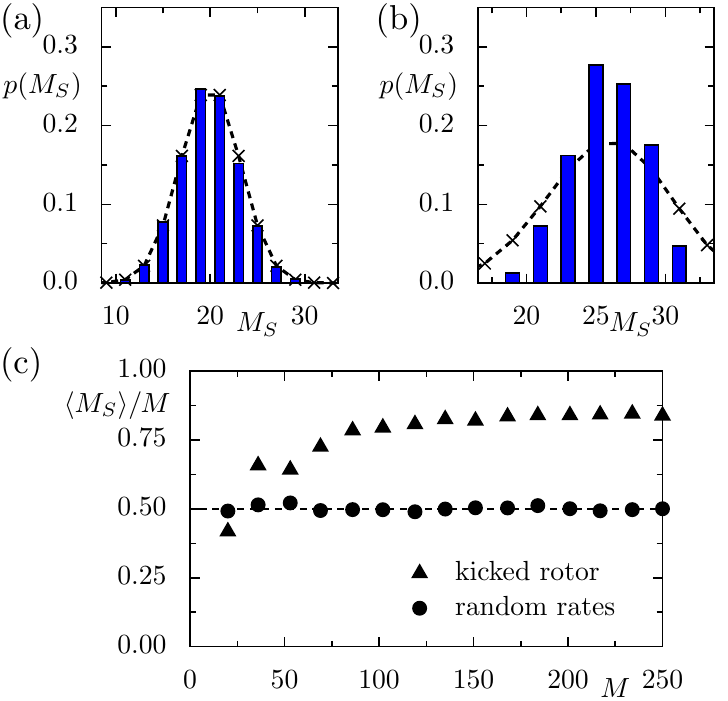}
	\caption{(a) Distribution of the number \(M_S\) of Bose-selected states for 5000 realizations
	of exponentially distributed random rates \(R_{ij}\) for a system of size \(M=40\). The
	distribution of the odd \(M_S\) is binomial with \(p=q=0.5\)
	(given by the black crosses connected by the dashed line). (b)~Same as in (a) but for 5000 realizations of a
	chaotic quantum kicked rotor with \(M=40\) Floquet modes where we choose the kicking strength K randomly
	from the interval \(K \in [9.5, 10.5)\). Binomial distribution with \(p=0.62\) for comparison. (c) Mean number \(\la M_S \ra\) divided by system size \(M\) of Bose-selected states
	as a function of \(M\) for the random rate model (dots) and the kicked rotor (triangles) both for 50 realizations of the system.
	For random rates \(\la M_S \ra\) coincides with the predicted value \(M/2\) (dashed line),
	however values for the kicked rotor deviate significantly from this result.}
	\label{fig:rotor-statistics}
\end{figure}

Figure \ref{fig:rotor-statistics}(a) shows the distribution of the number \(M_S\) of Bose-selected states
for random rates connecting \(M=40\) single particle states.
The odd number of Bose-selected states is given by a binomial distribution,
\begin{align}
	p(M_S) = \left\lbrace\begin{array}{cc}
	0 &\text{for } M_S \text{ even} \\
	\frac{1}{2^{M-1}}\left(\begin{array}{c} M \\ M_S \end{array}\right) &\text{for } M_S \text{ odd}
	\end{array}\right.
	\label{eq:distri-random}
\end{align}
centered around \(\la M_S \ra = M/2\) [cf.~Fig.~\ref{fig:rotor-statistics}(c)].
Such a behavior was observed in the literature
 for random rates,
mostly in the context of population dynamics, where similar equations, the Lotka-Volterra equations, appear \cite{FISHER199583, Chawanya02, Allesina05042011, VorbergEtAl13, knebel2015}.
Usually this result is motivated by arguing that for the random rates all states are equal, so that
 every state has the same probability to be Bose selected or not.
Taking into account the additional constraint that the total \(M_S\) must be odd, one can in this way motivate 
the distribution \eqref{eq:distri-random}
by just counting the number of possible choices of \(M_S\) states among the total \(M\) states.

However, the argument that every state has the same probability to be Bose selected or not must also follow
rigorously from the steady state equations, and thus also from Eq.~\eqref{eq:MS-problem}.
To fill this gap, we will compute the distribution of selected states.  
To this end, we will first contruct transformations of the rate-asymmetry matrix \(\matr{A} = (A_{ij})\) under which the number
of Bose-selected states \(M_S\) remains invariant, i.e.~the solutions of the problem~\refeq{eq:MS-problem} have the same \(M_S\).

\label{sec:random-rates}

\subsubsection{Reordering transformations}
Since the indexing of the states is arbitrary,
\(M_S\) will remain invariant when reordering the states.
The matrix \(\matr{T}_{k, l}\) describes the transpositions that exchange state \(k\) and \(l\).
Since \(\matr{T}_{k, l}^{-1} = \matr{T}_{k, l}\) the corresponding transformation takes the form
\begin{align}
	\matr{A} \rightarrow \matr{T}_{k, l} \, \matr{A} \, \matr{T}_{k, l}.
	\label{eq:Transposing-trafo}
\end{align}
Under this transformation, the matrix remains skew-symmetric,
and also the number of Bose-selected states remains invariant.

So for discussing the properties of a general \(\matr{A}\) with a given number \(M_S\) of Bose-selected states, we can
thus assume solutions of the form
\begin{equation}
{\eta} = (\eta_1, ..., \eta_{M_S}, 0, ..., 0)^T.
\label{eq:occup-ordered}
\end{equation}

\subsubsection{Rescaling transformations}
We are only interested in the number of selected states
and not in their occupation. Therefore, let us consider a rescaling of the coefficients \(\eta_i\)
and \(\mu_i\).
This is accomplished by a transformation induced by the matrix
\begin{align}
	\matr{D}_{\boldsymbol{\lambda}} = \mathrm{diag}(\lambda_1, ..., \lambda_M),
	 \quad \text{with positive} \ \lambda_k > 0.
\end{align}
As we want to apply this matrix to both the vectors \(\eta\) and \({\mu}\), it is important
that only positive rescaling $\lambda_k > 0$ is allowed. Otherwise we would transform a solution of problem \eqref{eq:MS-problem}
into vectors that do not solve a problem of this type.
Since the inverse of this matrix is again diagonal, \(\matr{D}_{\boldsymbol{\lambda}}^{-1} = \matr{D}_{(1/\lambda_1, ...)}\)
we find that the rescaling transformation
\begin{align}
	{\eta} \rightarrow \matr{D}_{\boldsymbol{\lambda}} \eta, \quad
	{\mu} \rightarrow  \matr{D}_{\boldsymbol{\lambda}}^{-1} {\mu}, \quad
	\matr{A} \rightarrow \matr{D}_{\boldsymbol{\lambda}}^{-1} \matr{A}  \matr{D}_{\boldsymbol{\lambda}}^{-1}
	\label{eq:diag-trafo}
\end{align}
preserves skew-symmetry of the matrix \(\matr{{A}}\). Note that we multiply \(\matr{A}\)
with the inverse from the left and the right. 
The rescaled quantities \(\tilde{\mu}, \tilde{\eta}, \matr{\tilde{A}}\) with \(\tilde{S}=S\) fulfil
again Eq.~\eqref{eq:MS-problem}, since
\begin{align}
	{\tilde{\mu}} = \matr{D}_{\boldsymbol{\lambda}}^{-1} {\mu} =
	 \matr{D}_{\boldsymbol{\lambda}}^{-1} \matr{A} \matr{D}_{\boldsymbol{\lambda}}^{-1} \
	 \matr{D}_{\boldsymbol{\lambda}} \eta = \matr{\tilde{A}} \tilde{\eta}.
\end{align}

\subsubsection{Standard matrices}
As we know that for any given skew-symmetric \(\matr{A}\) the problem \eqref{eq:MS-problem} has a unique solution \(S\) \cite{VorbergEtAl13,VorbergEtAl15},
we always find a sequence of transformations such that
\begin{align}
	{\tilde{\eta}}= (\underbrace{1, ..., 1,}_{M_S \text{ entries}} 0, ..., 0)^T ,
	\quad {\tilde{\mu}} = (\underbrace{0, ..., 0,}_{M_S \text{ entries}} -1, ..., -1)^T
	\label{eq:standard-form}
\end{align}
holds.
As the transformations \eqref{eq:Transposing-trafo} and \eqref{eq:diag-trafo} are invertible, it turns out that
it suffices to discuss the properties
of the set of standard matrices \(\matr{A}_{M_S}\) that is constructed such that it gives rise to a solution of the form \eqref{eq:standard-form}. 
Then the space of rate-asymmetry matrices
with selection number \(M_S\) is spanned by a sequence of the two physically intuitive transformations discussed above.

We find that these standard matrices take the form
\onecolumngrid
\begin{align}
	 \matr{A}_{M_S} =
\left(
\begin{array}{ccccccc} 
 & & & & x^{(1)}_{M_S+1} & \cdots & x^{(1)}_{M} \\
   & \matr{A}^S_{M_S}  & & & x^{(2)}_{M_S+1} -x^{(1)}_{M_S+1}& \cdots &x^{(2)}_{M} - x^{(1)}_{M} \\
      &  & & & x^{(3)}_{M_S+1} -x^{(2)}_{M_S+1}& \cdots &x^{(3)}_{M} - x^{(2)}_{M} \\
 & & & &\vdots & \ddots & \vdots\\
 & & & & 1- x^{(M_S)}_{M_S+1} & \cdots & 1-x^{(M_S)}_{M}\\
 - x^{(1)}_{M_S+1} & - x^{(2)}_{M_S+1} + x^{(1)}_{M_S+1} & \cdots  & -1+ x^{(M_S)}_{M_S+1}& \\
 - x^{(1)}_{M_S+2} & - x^{(2)}_{M_S+2} + x^{(1)}_{M_S+2} & \cdots  & -1+ x^{(M_S)}_{M_S+2}& \\
 \vdots  & \vdots  & \ddots  & \vdots & & \matr{A}_{\mathrm{arb}} \\
 - x^{(1)}_{M} & - x^{(2)}_{M} + x^{(1)}_{M} & \cdots  &- 1 + x^{(M_S)}_{M}& \
\end{array}
\right),
\label{eq:A-ms-general}
\end{align}
\twocolumngrid\noindent
with arbitrary real numbers \(x^{(i)}_j\), arbitrary (\(M-M_S\))-dimensional skew-symmetric matrix \(\matr{A}_{\mathrm{arb}} \)
and \(M_S\)-dimensional skew-symmetric matrix \(\matr{A}^S_{M_S}\) restricted by the existence of the homogeneous solution
\begin{equation}
	\matr{A}^S_{M_S} \quad \left( \begin{array}{c} 1 \\ \vdots \\ \ \\ 1 \end{array}\right) = 0.
\end{equation}
\subsubsection{Consequence for random rates}
For a random rate matrix, all entries \((A_{ij})_{j>i}\) of the rate asymmetry are random and statistically
independent.
We aim to find the probability to randomly choose
a matrix that is from the class generated by the \(\matr{A}_{M_S}\)-matrix
(under the transformations \eqref{eq:Transposing-trafo} and \eqref{eq:diag-trafo}).
To this end, we first count the number \(d\) of degrees of freedom in determining a general matrix \(\matr{A}_{M_S}\). 
In a second step we then discuss the influence of the transformations.

Let us begin with the block \(\matr{A}^S_{M_S}\).
To construct such a matrix, we start from an arbitrary \(M_S\)-dimensional skew-symmetric matrix 
which has
\(d_S = \frac{1}{2} \left(M_S -1 \right) M_S \)
degrees of freedom. 
We have to distinguish two different cases: For odd \(M_S\) this matrix has always a homogeneous solution,
for even \(M_S\) we have to fine tune one parameter for a homogeneous solution to exist, which reduces the number of
degrees of freedom by one.
Furthermore we have to subtract \(M_S-1\) degrees
because the homogeneous solution is pinned to \((1, \dots, 1)^T\).
Thus the number
of degrees of freedom in the subspace of the selected modes is
\begin{align}
	d_S^o&=\frac{1}{2} \left(M_S -1 \right) \, (M_S-2),\\
	d_S^e&=d_S^o- 1
\end{align}
for an odd or even number of selected modes, respectively.

Then, there are \(M - M_S\) rows, each of which has \(M_S-1\) free variables \(x^{(i)}\) to choose. This contributes
\begin{equation}
	d_f= (M-M_S)\, (M_S - 1)
\end{equation}
degrees of freedom.

Also there is still an arbitrary \((M-M_S)\)-dimensional skew-symmetric matrix free to choose, which adds another
\begin{equation}
	d_{\mathrm{arb}}= \frac{1}{2} (M - M_S) \, (M - M_S - 1)
\end{equation}
degrees of freedom.

This sums up to
\begin{align}
	d^{o} &= d^{o}_S + d_f + d_{\mathrm{arb}} = \frac{1}{2} (M-1) (M - 2) \\
	d^{e}  &= d^{o} - 1
\end{align}
for odd and
for even \(M_S\) respectively.
Interestingly for every \(M_S\) the number of free parameters of the generating matrix \(\matr{A}_{M_S}\) depends on the
parity of \(M_S\) only. Matrices with an even \(M_S\) have one degree of freedom less.

If we now choose a rate-asymmetry matrix randomly, the probability to
hit a matrix with a specific number of selected states is proportional to its size in parameter space.
The diagonal transformations \eqref{eq:diag-trafo} do not favour any number of selected states, as
they always contribute \(M\) degrees of freedom.
For an even number of selected states there is one free parameter less than for an odd number \(M_S\),
such that their generating matrices
form a set of probability measure zero in parameter space. Therefore
all generators \(\matr{A}_{M_S}\) of odd selection numbers have equal size in probability space and the even numbers are suppressed.

It is left to discuss the influence of the reordering transformations \eqref{eq:Transposing-trafo}.
They allow to distribute the \(M_S\) selected states over the \(M\) states.
The number of possible configurations for this is given by the binomial factor
\be
 	\left( \begin{array}{c} M \\ M_S \end{array} \right).
\ee
 After normalization, we infer
the distribution \eqref{eq:distri-random}.

\subsection{Chaotic quantum kicked rotor}
\label{sec:kicked-rotor}
We want to compare these results for uncorrelated random rates to chaotic
systems. 

A paradigm for quantum chaos is the quantum kicked rotor, a one-dimensional rotor
governed by the Hamiltonian
\be
	\hat{H}(\hat{\varphi},\hat{p},t) = \frac{\hat{p}^2}{2} + K \cos(\hat{\varphi}) \sum_{n\in Z} \delta(t-n)
\ee
with time-periodic kicks of strength \(K\), period $\tau=1$ and $\left[ \hat{\varphi},\hat{p} \right] = \mathrm{i} \hbar_{\mathrm{eff}}$.
For \(\hbar_{\mathrm{eff}} = \frac{2\pi}{M}\), $M \in \mathbb{N}$, we can restrict the system to a torus with periodic coordinate \(\varphi \in [0, 2\pi)\) and periodic momentum \(p \in [-\pi, \pi)\).
Note that since the available phase space volume on the torus  is $V = (2\pi)^2$, there exist
 $V/(2\pi\hbar_\mathrm{eff}) =M$
Floquet states on the torus.

These Floquet states \(\ket{i(t)}\) are eigenstates of the one-cycle evolution operator
\begin{align}
	\hat{U}(1, 0) = \mathrm{e}^{-\frac{\mathrm{i}}{\hbar_{\mathrm{eff}}} K \cos(\hat{\varphi})} \  \mathrm{e}^{-\frac{\mathrm{i}}{\hbar_{\mathrm{eff}}} \frac{\hat{p}^2}{2}}
\end{align}
fulfilling
\(\hat{U}(1, 0) \ket{i(0)} = \exp(-\mathrm{i} \varepsilon_i /\hbar_{\mathrm{eff}})\ket{i(0)}\) with corresponding quasienergies \(\varepsilon_i\).

This kicked rotor is coupled to a bath with temperature~\(T\).
We consider the coupling operator
\be
\vo= \sin(\hat\varphi) +\cos(\hat\varphi), 
\ee
which respects the periodicity of \(\varphi\) and breaks the parity (such that also even and odd Floquet states are coupled to each other).

In Fig.~\ref{fig:rotor-statistics}(b) we show the distribution for the number of selected states  \(M_S\)
that we obtain when randomly choosing the
kicking strength \(K\) within the interval \([9.5, 10.5]\) (where the classical counterpart of the quantum kicked rotor is essentially fully chaotic)
 for a rotor with \(M=40\) Floquet states.
From Fig.~\ref{fig:rotor-statistics}(b) and (c) it is clear that the random rate
model fails to predict the number of selected states for a typical realization of the chaotic quantum kicked rotor.
The distribution is centered around a much larger value than \(M_S=M/2\) expected for random rates [Fig.~\ref{fig:rotor-statistics}(a)].
It also seems that the distribution may not be fitted with a binomial distribution, which for example for \(p=0.62\) (black crosses)
is much broader then the one that is observed.
This trend also manifests itself in Fig.~\ref{fig:rotor-statistics}(c) where the triangles
show the mean number \(\la M_S \ra\) of Bose-selected states for the quantum kicked rotor.
This number lies well above \(M/2\), for systems of size \(M \gtrsim 150\) about 80\% of the states are Bose selected.
Consequently, we come to the intriguing conclusion that
there must be additional correlations among the rates \(R_{ij}\) that are responsible for the fact that significantly
 more states are selected for the chaotic quantum kicked rotor than in the random-rate model. 
In the Appendix, we describe a model with correlated random rates that shows a similar distribution of selected states as 
the quantum kicked rotor model. However, the origin of the large number of selected states 
for our quantum kicked rotor model remains an interesting open question. 

\section{Rates with product structure}



Consider an arbitrary time-independent system with Hamiltonian \(H_S\) which is coupled to a positive temperature bath (\(B_1\))
and a population-inverted bath (\(B_2\)) described by a negative temperature
through coupling operators which obey the form of a projector on a single quantum state, 
\begin{equation}
	\vo^{(B_1)} = \ket{f}\bra{f}, \qquad \vo^{(B_2)} = \ket{g}\bra{g}.
	\label{eq:prod-coupling}
\end{equation}
Note that the states $\ket{f}$ and $\ket{g}$ can also be coherent superpositions of the single-particle eigenstates $i$.
In this section we show that the number of selected states remains always smaller or equal than three, \(M_S \leq 3\),
 for all system sizes~\(M\). 
Note that in case of this specific system-bath coupling even a system with
chaotic single--particle dynamics features only a maximum number of three condensates.
An example of a system with such a system bath coupling is the one depicted in Fig.~\ref{fig:main}(b).
Here the states $f$ and $g$ correspond to the local Wannier orbitals at lattice sites $\ell_1$ and $\ell_2$, respectively. 

For coupling operators of the form \eqref{eq:prod-coupling}, we  find the rate asymmetry matrix
\begin{align}
	A_{ij} &= \frac{2\pi}{\hbar} J(\varepsilon_j-\varepsilon_i) (f_i f_j - g_i g_j)
\end{align}
 from Eq.~\eqref{eq:asym-1b}, where \(f_i = \left\vert \braket{i}{f}\right\vert^2\), and \(g_i = \left\vert \braket{i}{g}\right\vert^2\).
Here we assume that \(J_{B_2}(\varepsilon) = -J_{B_1}(\varepsilon) = -J(\varepsilon)\),
giving rise to rates \(R_{ij} \geq 0\).
Moreover, let us first consider ohmic baths with spectral density \(J(\varepsilon) \propto \varepsilon\). We find
a rate asymmetry matrix having the product struture
\begin{align}
	A_{ij} &\propto f_i f_j\varepsilon_j - f_i f_j\varepsilon_i - g_i g_j \varepsilon_j + g_i g_j \varepsilon_i.
\end{align}

Now let \(\eta_{i}\) be the solution of Eq.~\eqref{eq:MS-problem}. 
It then follows that in the subspace of selected states one has
\begin{equation}
	\begin{split}
	0= \mathbf{A} \eta=&  \left(\begin{array}{c}
		f_{i_1}\\
		f_{i_2}\\
		...\\
		f_{i_{M_S}}\\
	\end{array}\right) c_1
	- \left(\begin{array}{c}
		f_{i_1} \varepsilon_{i_1}\\
		f_{i_2} \varepsilon_{i_2}\\
		...\\
		f_{i_{M_S}} \varepsilon_{i_{M_S}}\\
	\end{array}\right) c_2\\
	&-
	\left(\begin{array}{c}
		g_{i_1}\\
		g_{i_2}\\
		...\\
		g_{i_{M_S}}\\
	\end{array}\right) c_3
	+ \left(\begin{array}{c}
		g_{i_1} \varepsilon_{i_1}\\
		g_{i_2} \varepsilon_{i_2}\\
		...\\
		g_{i_{M_S}} \varepsilon_{i_{M_S}}\\
	\end{array}\right) c_4
	\end{split}
	\label{eq:vector-chain}
\end{equation}
with
\begin{equation}
	\begin{split}
  		c_1 = \sum_{i \in S} \varepsilon_i f_i \eta_{i},\quad
  		c_2 = \sum_{i \in S} f_i \eta_{i},\\
  		c_3 = \sum_{i \in S} \varepsilon_i g_i \eta_{i},\quad
  		c_4 = \sum_{i \in S} g_i \eta_{i}.
	 \end{split}
\end{equation}
Since \(f_i >0\), \(g_i >0\) and \(\varepsilon_i > 0\) (otherwise we can always
shift all \(\varepsilon_i\) by some constant), these coefficients are positive, \be
c_i > 0.
\ee

Now if \(M_S =1\) or \(M_S = 3\) then the four vectors in Equation \eqref{eq:vector-chain} will be linearly dependent,
thus the \(c_i\) can be positive as they should.
However, if \(M_S\) was greater than three, then generally the four vectors will be linearly independent, so that \(c_i=0\) must hold,
in contradiction to the assumption \(c_i > 0\).

Therefore we have shown that for two ohmic baths with product coupling
to the system, the number of Bose-selected states is restricted to a maximum of three.

\begin{figure}[t]
	\includegraphics{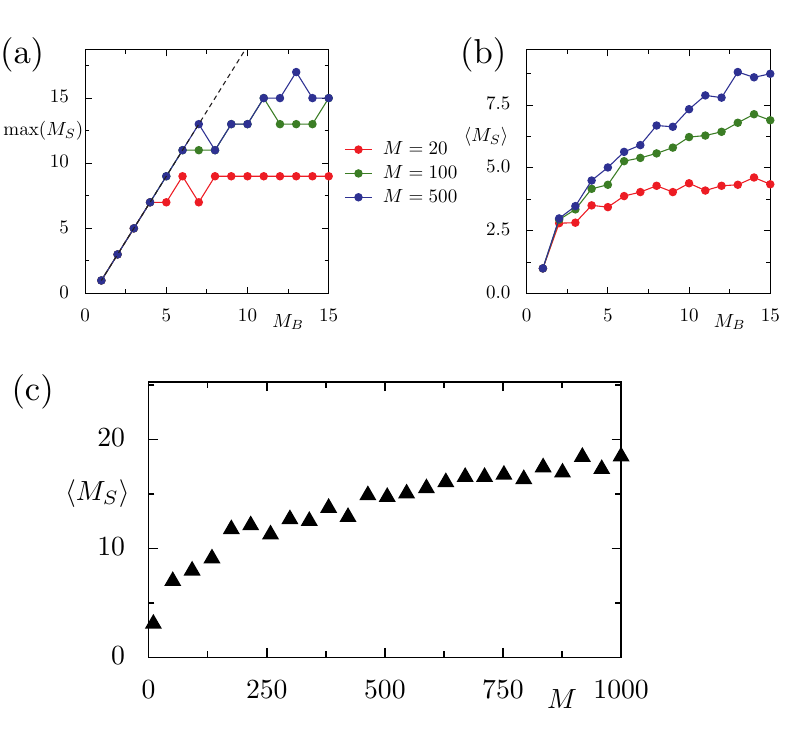}
	\caption{ (a) Maximum number and (b) mean number \(M_S\) of Bose-selected states
	observed for \(200\) realizations of a random chaotic
	system \(H_S\in \mathrm{GOE}(M)\) as a function of the number of baths \(M_B\) coupled to the system.
	We place \(M_B\) ohmic baths at some random index \(i\) (coupling operator \(\vo = \ket{i}\bra{i}\)).
	We choose half of the baths with positive, the other half with negative temperature (for odd \(M_B\) we randomly decide).
	The dashed line in (a) is the predicted upper bound \(2M_B - 1\). (c) mean number \(M_S\) of selected states in this model
	(here only 50 realizations are used)
	but as a function of system size \(M\). We choose \(M_B= M/2\), i.e.~we are in the regime of so many baths that the number \(M_S\)
	is saturated. 
	}
	\label{fig:M_S-randmatr}
\end{figure}

There is a straight-forward generalization of this simple algebraic argument
to the case where \(M_B\) ohmic baths are coupled to an autonomous system
via coupling operators of the product form, Eq.~\eqref{eq:prod-coupling}.
In this case, we find an analog to Eq.~\eqref{eq:vector-chain}, but with
a linear combination of \(2 M_B\) vectors.
By similar reasoning the number of selected states is then restricted to
\(M_S \leq 2 M_B -1\). 

We expect that there is a generalization of the above arguement also to non-ohmic systems described by arbitrary spectral densities. 
 We have checked functions of the form (remember that $J$ must be odd)
\be
	J(\varepsilon) \propto \vert \varepsilon  \vert^d \mathrm{sgn}(\varepsilon)
\ee
with some power $d$ (not necessarily integer) and similarly observe \(M_S \leq 2 M_B -1\).

Fig.~\ref{fig:M_S-randmatr}(a) confirms the result for ohmic spectral densities $J(\varepsilon) \propto \varepsilon$. It shows the maximum
number \(M_S\) of selected states for 200 systems that are randomly drawn from the
gaussian orthogonal ensemble, GOE\((M)\), the ensemble of orthorgonal \(M \times M\) matrices, where
the probability to find a matrix \(H\) is given by \(p(H) \propto \exp(-\frac{M}{4}\mathrm{tr}(H^2))\).
In random matrix theory these Hamiltonians serve as a model for
a fully chaotic system with time-reversal symmetry.
We choose random Hamiltonians to make sure that the number of Bose-selected states
is not additionally restricted by the system dynamics.
We couple these systems to \(M_B\) ohmic baths, where we randomly choose
an index \(i \in\lbrace1,\dots,M\rbrace\) to which the bath is coupled with operator \(\vo = \ket{i}\bra{i}\).
For half of the baths we choose positive temperature, for the other half negative temperature
(the selected states are only determined by the rate asymmetry matrix $A$ and thus independent 
of the absolute values of these temperatures).

In Fig.~\ref{fig:M_S-randmatr}(a) we plot the maximum number of selected states found for an ensemble of 
$200$ realizations of this model. For sufficiently small $M_B$ it equals the predicted upper bound for $M_S$. 
However, we observe that for larger values of \(M_B\), the observed maximum number of selected states
saturates very quickly at values that are of the order of \(\sqrt{M}\) where \(M\) is the system size.
This bounds the mean number of selected states as shown in Fig.~\ref{fig:M_S-randmatr}(b).

We would like to stress that the behavior of the autonomous GOE
is very different from that of the time-periodically driven rotor, although
both systems exhibit chaotic single particle dynamics.
First, in the limit of large system size, \(M\rightarrow \infty\), the number of
selected states can be intensive, as long as the number of baths \(M_B\) is not
scaled with system size,
so that we may find fragmented condensation in the thermodynamic limit.
However, even if we scale \(M_B\) with system size, as shown in Fig.~\ref{fig:M_S-randmatr}(c) for \(M_B = M/2\),
the mean number of selected states seems to scale strongly sublinear with at maximum \(\la M_S \ra \propto \sqrt{M}\),
rather than the drastically different extensive \(\la M_S \ra \propto {M}\) scaling that is observed for time-periodically driven systems,
cf. Sec.~\ref{sec:floquet-chaotic}.

\section{Conclusion and outlook}
In this paper, we have addressed the effect of Bose selection in nonequilibrium steady states of driven-dissipative ideal Bose gases that exchange energy with an  environment they are weakly coupled to. Namely, when the total particle number $N$ is increased in such a system, eventually only the occupation of a number $M_S$ of selected modes will increase linearly with $N$, while the occupation of all other modes saturates. So far, only very little was known about the factors that determine how many modes will be selected. 

Within this paper, we have investigated the question of how many Bose-selected states will be found by addressing different relevant scenarios leading to different forms of the rate matrix describing the driven-dissipative ideal Bose gas. We have shown that upper bounds for the number of Bose-selected states $M_S$ that are independent of the dimensionality $M$ of the underlying single-particle state space appear both for systems whose rates can be understood as the discretization of a continuous function as well as for systems  that couple to baths via a few single-particle quantum states only. In these cases Bose selection corresponds to (fragmented) Bose condensation, since each Bose-selected state acquires a macroscopic occupation in the thermodynamic limit $M\to\infty$ at fixed density $N/M$. 

Moreover, we have discussed two scenarios where the number of selected states is found to grow linearly with $M$, so that none of the selected states will acquire a macroscopic occupation in the thermodynamics limit. On the one hand, we have shown that for randomly drawn uncorrelated rates the number of selected states follows a binomial distribution so that on average half of the single-particle states are selected. On the other hand, we have numerically treated a periodically driven system, the quantum kicked rotor, in a regime where the corresponding classical model is fully chaotic. Averaging over various kicking strengths we find that more than half (about 75 percent) of the states become selected. This is an intriguing result. It implies that the chaotic driven system must give rise to correlations among the rates, which are responsible for a significant enhancement of the selected states. While we were able to construct a model of correlated random rates showing similar behavior, the question about the origin and the nature of the large number of Bose-selected states for the quantum kicked rotor remains open. 

Further open questions to be addressed in future research
 include the role of interactions and larger system bath coupling on the effect of Bose selection, 
 an understanding of the factors determining the number of Bose-selected states in photonic systems where 
 Bose selection is induced by the interplay of pumping, particle loss, 
 and thermalization \cite{VorbergEtAl2018}, and the investigation of possible experimental platforms. 

\begin{acknowledgments}
We acknowledge discussions with Daniel Vorberg.
This work was supported by the German Research Foundation
DFG via the Research Unit FOR 2414.
\end{acknowledgments}

\appendix*
\section{Modified random wave model
for the rates in the quantum kicked rotor}

\begin{figure}[t]
	\includegraphics{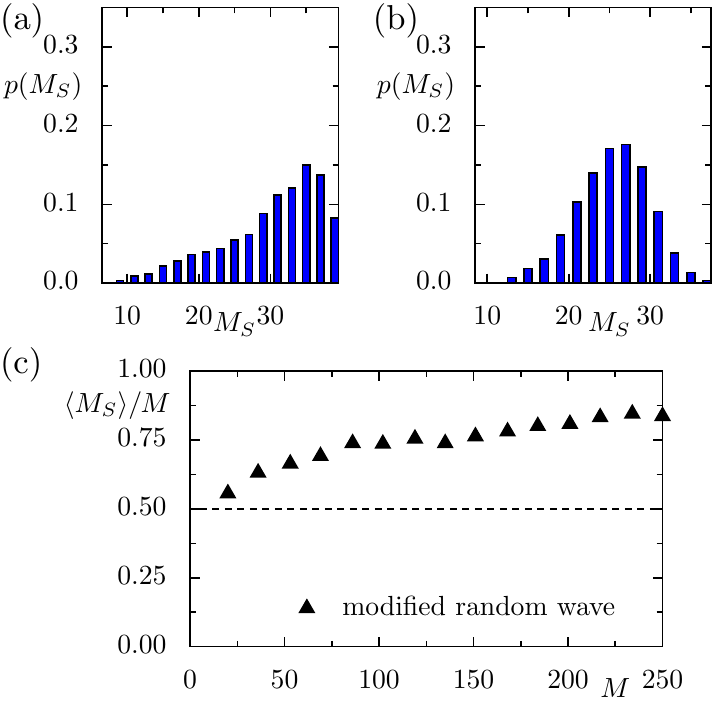}
	\caption{(a) Distribution of the number \(M_S\) of Bose-selected states for 5000 realizations
	of rates \(R_{ij}\) stemming from the random wave model, Eq.~\eqref{eq:rand-wave},
	 with \(\kappa = 1.9 \pi M\), \(L=80\) and \(M=40\) states.
	 (b)~Same as in (a) but for 5000 realizations of the modified random wave model, Eq.~\refeq{eq:rand-wave-mod},
	 with \(M=40\) states, \(\kappa = 1.9 \pi M\), \(L=80\) and \(\lambda=10\).
	 (c) Mean number \(\la M_S \ra\) of Bose-selected states
	 divided by system size \(M\) as a function \(M\) for the modified random wave model (triangles) for 50 realizations of the system.
	We use \(L=2M\) components.
	The dashed line is at \(M_S = M/2\).}
	\label{fig:random-wave-statistics}
\end{figure}

The deviations from the random rate model indicate that the rates
of the chaotic quantum kicked rotor contain correlations that lead to the Bose selection
of more states than in the uncorrelated case.
In this appendix we construct a random rate matrix having correlations that lead to similar behavior. 

Correlations of the rates associated with the quantum-kicked-rotor model in contact with a heat bath that we considered
were discussed in Ref.~\cite{Wustmann10} (where the single-particle problem was studied).
The rate
\(R_{ij}\) and the backwards rate were proposed to obey
\be
	R_{ji} = (1 + \xi_{ij}) R_{ij}, \text{ for }i>j,
	\label{eq:rate-model-wust}
\ee
with both \(R_{ij}\) and \(\xi_{ij}\) stemming from individual exponential distributions
with scale parameters \(\lambda\) and \(\lambda_\xi\), where \(\lambda_\xi\) decreases with system size
as \(\lambda_\xi \propto M^{-1.2}\).
However, the rate model \eqref{eq:rate-model-wust} leads to steady states with even less then half of the states being
Bose selected (data not shown).

Note that in Sec.~\ref{sec:M_s-func-k} we have encountered a model, which also features that typically
more than half of the states are Bose selected. It is the random wave model, Eq.~\eqref{eq:rand-wave},
with parameter \(\kappa \approx 1.9 \pi M\) [cf.~also Fig.~\ref{fig:rand-wave-func-k}].
A suggestive point of view for why the random wave model might be suitable to approximate rates of a chaotic
map, is that random waves have been used successfully to model typical chaotic eigenstates \cite{berry1977, Loeck2010, Berry02}.
Note, however, that this vague reasoning is not based on a microscopic picture for the derivation of the rates. 

We can see  in Fig.~\ref{fig:random-wave-statistics}(a) that the distribution for the random wave model with \(M=40\)
is much broader than the one we observe for the quantum kicked rotor in Fig.~\ref{fig:rotor-statistics}(b).
Also the rates \(R_{ij}\) that result from a random wave model, Eq.~\eqref{eq:rand-wave}, do not follow an
exponential distribution. Their distribution \(p(R_{ij})\) is rather peaked at some finite value.
To correct for this, we introduce a modified random wave model for the rates
\be
	R(k, q) = \sum^L_{l=1}\ \vert c_l \vert \mathrm{Re}\big\lbrace 1+
	\mathrm{e}^{\mathrm{i} (\kappa_{k,l} k + \kappa_{q,l} q + \alpha_l)}\big\rbrace
	\mathrm{e}^{-\lambda \vert k-q\vert},
	\label{eq:rand-wave-mod}
\ee
where we constrain the waves to positive values and localize them on a length \(\lambda^{-1}\) by introducing an exponential factor. 
In this model we choose \(\kappa_{k,l} = \kappa \sin(\varphi_l)\), \(\kappa_{q,l} = \kappa \cos(\varphi_l)\)
similar to the random wave model with fixed absolute value of the
wavenumber \(\kappa\), uniformly distributed angles \(\varphi_l , \alpha_l\in \left[0, 2 \pi \right]\) and \(c_l\) from a normal distribution.

We observe that the modified random-wave model has
rates that are distributed exponentially and the distribution of the \(M_S\) shows relatively
 good agreement with the data from the quantum kicked rotor for \(\kappa = 1.9 \pi M\) and localization parameter \(\lambda=400/M\).
This can be seen for example by comparing the distribution for \(M=40\) discrete states in Fig.~\ref{fig:random-wave-statistics}(b)
to the distribution of the quantum kicked rotor in Fig.~\ref{fig:rotor-statistics}(b),
although the mean values coincide, the distribution of our model is, however, a bit broader than the one obtained for the quantum kicked rotor.
Also as a function of system size, Fig.~\ref{fig:random-wave-statistics}(c), the model seems to reproduce
the mean value of selected states in Fig.~\ref{fig:rotor-statistics}(c) quite nicely.

However, despite the fact that the modified random wave model \eqref{eq:rand-wave-mod} gives rise to a similar distribution 
of the number of selected states as the one obtained for the quantum kicked rotor, we have
 no evidence that the modified random-wave model mimics the physics of the quantum kicked rotor coupled to a heat bath 
 [see Sec.~\ref{sec:kicked-rotor}].

\bibliography{mybib}

\end{document}